\documentclass[11pt]{article}
\usepackage{graphicx}
\def\title{\begin{center}\Large\bf}
\def\author(s){\vspace{0.3cm}\large\rm}
\def\text{\end{center}}

\begin{document}


\noindent {\small{\it Kinematics and Physics of Celestial Bodies,
Suppl., No 4, 2003, pp. 211-216}}

\medskip \hrule \medskip

\bigskip
\bigskip


\title
Singular Sources of Energy in Stars and Planets

\bigskip



\author(s)
B.E. Zhilyaev

\bigskip

\smallskip

\noindent {\small {\it Main Astronomical Observatory, NAS of
Ukraine,  27 Zabolotnoho, 03680 Kiev, Ukraine}} \\

\noindent {\small {\it e-mail:}} {\small{\bf zhilyaev@mao.kiev.ua}}

\smallskip



\text


\section*{Abstract}

If primordial low-mass black holes (PBH) exist in the Universe than
many of stars and planetary bodies appear to be "infected" by them.
This is also true in regard to the Sun and likely Jupiter and
Saturn. The availability of even the very low-mass inner
relativistic reactor may lead to essential changes in evolution
scenario of a celestial body on its lifetime scale. Black holes in
stellar interior may be found either in consequence of captures
process or incorporation during the formation of a star from
interstellar clouds. Surprisingly that in the equilibrium state a
PBH growth is a long-lived process with e-folding rise time of
billion years. One can envision a PBH orbiting inside the Sun. Our
considerations showed that the PBH experiences very little friction
in passing through the stellar matter. If the BH mass is above
$10^{-5}M_{\odot}$  the major contribution to the luminosity comes
from the relativistic gravitational reactor. In such a case a star
evolves towards the Eddington limit. This should lead to
considerable expansion of a star and a global stability loss.
Microscopic PBHs can exist in the interior of planetary bodies too.
To produce the required excess of thermal energy on Jupiter and
Saturn the masses of PBH captured are assumed to be reached of
$4\cdot 10^{19}$  and $7\cdot 10^{18}$ g , respectively. These
microscopic objects are comparable to the hydrogen atom in size. One
can envision even a planet with the PBH acting as the
self-sufficient source of heating. Such a planet does not need a sun
to maintain animal life on its surface. This may last eons.

\vspace{1.0cm}

\noindent {{\bf keywords}$\,\,\,\,$\large black hole physics --
(stars:) planetary systems -- Solar system: general}

\large

\section{BASIC CONCEPT}

Is it the case that a small black hole captured inside a star causes
its collapse? That is not the case, otherwise, if such a situation
would be present that should produce observable signatures in the
form of supernova explosions at a fantastic rate as will be seen
below.

Hawking (1971) first pointed out that gravitationally collapsed
objects, which were formed in the early Universe, could have
accumulated at the center of a star like the Sun. The presence of
such a hypothetical situation requires a more detailed
consideration.

Following Beckenstein (1974) and Sciama (1976), one can consider the
BH in thermostat as a heat engine with an available thermodynamic
efficiency of $(2-\sqrt{2})=0.59$. Such a relativistic reactor may
be competitive in power with the nuclear burning. Here we attempt to
answer the questions: (1) where are BHs from in stellar interior?
(2) how does this feature change the stellar structure and
evolution? (3) may one expect some observable signatures of this
feature in the Universe?


\section{CAPTURE AND INCORPORATION}
\noindent The most important problem we are interested in is the
manner in which stars with singular source of energy arise. Black
holes in stellar interior may be found either in consequence of
captures processes or incorporation during the formation of a star
from interstellar clouds. We will refer to these  as "captured" or
"incorporated" BHs. First we consider the BH capture process, i.e.
the collisions between field stars and primordial black holes
(henceforth PBH).  In this way the BH moves down through the stellar
atmosphere and loses its kinetic and orbital angular momenta.
Eventually the BH may be captured. A critical question is the
estimation the proportion of stars invaded a BH. Consider collisions
between a star and a BH. Assume that the direction of motion of the
BH is distributed isotropic. Then the number of collisions is given
by
\begin{equation}\label{numcol}
N\approx \sqrt{2}\cdot n_{BH} <v_{BH}> \sigma \Delta t,
\end{equation} where $n_{BH}$ is the BH number density, $<v_{BH}>$
is the average velocity, $\sigma$ is the capture cross-section and
$\Delta t$ is the time interval of interest. The star-BH
cross-section is taken approximately equal to the star's
cross-section. The determination of the PBH number density is the
most vague problem. The credible source of PBHs is usually
identified with the earliest stages of the Universe. This is
undoubtedly one of the most discussing topics of cutting-edge
cosmology. As discussed by many authors (Fukuda et al. 1998, Turner
1999), only about one-tenth of the baryons are visible in the form
of bright stars.  At very large red shifts, according to the latest
data on Light Echoes of radio sources, the baryonic density of the
Universe is still lower and comes to only $\sim 0.001$ of the total
(Sholomitskii 1997). From the empirical standpoint, this may support
the argument that some fraction of dark matter in the Universe
exists in the form of PBHs. Primordial black holes either formed
naturally, as being produced by the collapse during the radiation
era before $\sim 10^{-4}\,\mathrm{s}$ since the beginning of the
Universe, when the density was greater than nuclear density, or fed
into the Universe $ab$ $initio$, as discussed by Lin et al. (1976).
They have also noted that primordial black holes could grow up to
the horizon mass of about $1\: M_{\odot}$ at $10^{-4}\,\mathrm{s}$.
On the other hand, PBHs of less then $\sim 10^{15}\,\mathrm{g}$
should have evaporated by now through the Hawking process. This
introduces a low-mass cut-off in the spectrum. According to Carr
(1976), the present number density of PBHs for initial mass around
$m_{0}\simeq 10^{15}\,\mathrm{g}$ should be
\begin{equation}\label{powdensity}
n(m)=n_{0}\left(\frac{m}{m_{0}}\right)^{-k}
\end{equation}
where the exponent $k$ lies in the range $2< k <3$, depending on the
equation of state in the early Universe. These PBHs had formed at
time around $10^{-23}\,\mathrm{s}$. There is some uncertainty about
the exponent $k$ because of uncertainties involved in what model one
adopts for the early Universe. It appears that PBHs with mass of
$\sim 10^{15}\,\mathrm{g}$ are the most plentiful, irrespective of
the model. If we will require only an approximate solution of the
problem we may assume that all the PBHs are of some single mass,
$M$, i.e. $n(m)$ $ \sim $ $ \delta (m-M)$. Assume that PBHs amount
to $\alpha$ fraction of the total mass in the Galaxy. Using the data
given by Allen (1973) on the density in the solar vicinity
($\rho_{0} \simeq 6\,10^{-24}\,\mathrm{g\, cm^{-3}}$) and the
$\delta (m-m_{0})$ approximation for the mass spectrum, we find the
PBH density,
\begin{equation}\label{BHdensity}
n_{BH}=\frac{\alpha \:\rho_{0}}{m_{0}} \simeq 6\,10^{-39}\;
\alpha,\;\mathrm{cm^{-3}}
\end{equation}
Adopting the value of $<v_{BH}>$ = $100\,\,\mathrm{km\, s^{-1}}$ and
$\Delta t$ = $5\,\mathrm{Gyr}$ for the Sun lifetime we find the
number of BH collision events,
\begin{equation}\label{Number}
N \simeq 2\,10^{9}\, \alpha \end{equation} Even though the total
density of PBHs is by nine orders of magnitude less than that in the
Galaxy we may deduce that most of stars expected to be "infected" by
PBHs during their lifetime. Unlike the solid body, a BH would suffer
very little friction in passing through the stellar material. This
distinctive feature of BH was first noticed by Hawking (1971). To
illustrate this it is sufficient to consider the BH interaction with
the ambient stellar matter as the local straight-line inelastic
impact. Using the equation for the conservation of momentum, we may
write
\begin{equation}\label{impact}
v_{1}=v_{0}\frac{m_{BH}}{m_{BH}+m}
\end{equation}
where $v_{0}$, $ v_{1}$ denote the pre- and post-impact velocities,
$m$ is the total atmospheric mass which undergo the impact, $m\simeq
\pi\,R_{BH}^{2}<\rho>2\,R_\mathrm{star}\,\,\,,\,\mathrm{g}\,$. The
mass is restricted to a column with the BH in cross-section $\times$
the stellar diameter in length. We obtain
\begin{equation}\label{impact2}
v_{1}\simeq v_{0}(1-\frac{m}{m_{BH}})
\end{equation}
For the Sun and the most of stars the quantity $m/m_{BH}$ is
vanishing small ($\sim 10^{-17}$). So the BH passes practically
unobstructed through the body of a star, as through a vacuum. This
means that the capture process can take place involving only three
bodies, for example, a star and one of its planets. In such a case
the BH may be in an orbit deep inside star, over billion years,
until it is brought to rest at its center.

In the scenario of "incorporated" PBH its mass should be
proportional to some degree to the fraction of the PBH matter in the
density of baryons in the Universe. Even though the proportion of
PBHs is vanishing small ( $\alpha \sim 10^{-16}$ ), the absorption
probability value of a microscopic PBH by a protostar keeps close to
one.

The incorporation of PBHs during the formation of stars and other
gravitationally bound objects was analyzed by Derishev \& Belyanin
(1999). The detailed description of a gravitational incorporation
requires exact calculations of the collapse dynamics. Two of the
simplest cases were analyzed. These authors argued that in the
free-fall contraction relationship between the PBH density and the
average one remains constant. PBHs become trapped out inside a
protostar. In the case of an adiabatic contraction an appreciable
fraction of PBHs forms the gravitationally captured haloes around
the protostar. On the whole, from these calculations we can draw the
conclusion that some fraction of PBHs appears to be trapped out
inside a star during the contraction of protostellar clouds.

An estimate of the space density of primordial black holes can be
obtained from the flux of the diffuse extragalactic $\gamma$-rays
(Chapline 1975, Page and Hawking 1976, Lin et al. 1976, MacGibbon
and Carr 1991). This radiation was produced in the
quantum-mechanical decay of the low-massest PBHs created in the
early Universe. The moment of PBH formation $t_{0}$  depends on its
starting mass (Zeldovich and Novikov 1966), $t_{0}(\mathrm{s})\simeq
GM/c^{3}\simeq 2\,10^{-39}\,M(\mathrm{g})$. The hypothesis of PBHs
formation near the cosmological singularity from density and metric
fluctuations validated through numerical calculations by Novikov et
al. (1979).

Observations place an upper limit on an average space density of
PBHs about of $10^{4}\,\mathrm{pc^{-3}}$. But if  PBHs are clustered
into galaxies, the local density can be greater by a factor
exceeding $\sim 10^{6}$ (Page and Hawking 1976, Wright 1996). This
provides an upper limit  of about $n_{BH} \sim 4
\,10^{-46}\,\mathrm{cm^{-3}}$ in the Galaxy (Chapline 1975, Wright
1996). Observations of the Hawking radiation from the globular
clusters can provide next observational signature of PBHs.
Gravitationally captured PBH haloes around the globular clusters
were considered by Derishev and Belyanin (1999). EGRET observations
of the $\gamma$-ray luminosity above $100\,\mathrm{Mev}$ of five
nearby massive globular clusters placed, however, only the upper
limits on the total mass of PBHs and their mass ratio in these
clusters ($\alpha \leq 2\,10^{-6}$).

\section{LONG LIFE}

\noindent Let us suppose that a small BH exists in the center of a
star and explore the consequences stemmed from this hypothesis. We
may assume that a BH was either captured by star from space or
incorporated during the process of star formation or occurred for
some another reason. The advanced theoretical treatment of both
thermo- and gas-dynamical structure of a BH in dense thermostat is
that to be still investigated in detail. Now, we are interested in
the overall picture of the problem. In the ambient BH atmosphere the
radiation pressure is considered to balance the gravity in the
equilibrium state. The radiation pressure arises because of the
accretion of gas onto the BH when some part $\beta$ of the original
rest-mass of particles is radiated away. So, in treating the problem
the Eddington solution (Eddington 1926) is appropriate to use with a
sufficient degree of accuracy. For hydrogen plasma with the Thomson
scattering the BH luminosity should be no greater than Eddington's
limit
\begin{equation}\label{Eddlim}
L_{c}=\frac{4\pi G m_\mathrm{p} c M}{\sigma_\mathrm{T}}= 3\,10^{4}
L_{\odot}\frac{M}{M_{\odot}},
\end{equation} where G is the gravitational constant, $m_\mathrm{p}$ the proton mass, $c$ the speed of light, $\sigma_\mathrm{T}$  the Thomson cross-section and $M$ the BH mass. At low BH mass it may be thought that significant fluxes of photons with energies greater than $\sim \mathrm{1\, MeV}$ could be produced. In this case the scattering cross-section in (7) must be replaced by the cross-section for Compton scattering. If the accretion rate sets the pace for energy release at a level of Eddington's limit, the BH mass growth rate is given by equation (Zeldovich \& Novikov 1971)
\begin{equation}\label{growth}
-\varphi (r) \frac{d M}{d t}=\frac{G M}{R}\frac{d M}{d t}=L_{c},
\end{equation}
where $ -\varphi (r)=\beta c^{2}$ is the effective gravitational
potential. For a Schwarzschild and a Kerr black hole $\beta$ equals
about 6\% and 42\%, respectively (Petropoulos \& Mavrogiannis 1995).
The BH mass growth rate may be written as
\begin{equation}\label{growth2}
\frac{1}{M}\frac{d M}{d t}=\frac{4 \pi G m_\mathrm{p}}{\beta c
\sigma_\mathrm{T}}=7\,10^{-17}\beta^{-1}\mathrm{s^{-1}}=(0.5
\;\beta\; \mathrm{Gyr})^{-1}
\end{equation}
Surprisingly that in the equilibrium state BH growth is a long-lived
process with e-folding rise time of billion years. As expected,
specific boundary conditions have to stabilize the neighborhood of
the BH. The stability arises because the gravitational reactor is
immersed into a huge reservoir of dense heat-retaining matter, of
the kind of "prison  stability". The behavior of an unstable mode
can be investigated if it is viewed as oscillator. Strong restoring
forces are supplied by the radiation pressure with increasing the
accretion rate and by the gravity  with its decreasing. One of the
characteristic frequencies governing the dynamic stability of the BH
near environment is the Lamb frequency (Unno et al. 1979), given by
the local sound speed divided by the BH size. Due to its
dramatically high value of about tens of MHz compared with that of
the solar core p- and g-modes, these instabilities it seems to be
damped. It is pertinent to cite one illustrative example. The
contribution of the PBH of $M=10^{-5}M_{\odot}$ to the total solar
luminosity is estimated to be 30\% according to (7). The BH size is
$R =2 G M/c^{2} \simeq 1\,\mathrm{cm}$, its effective temperature
$T_\mathrm{eff}\simeq 10^{8}\,\mathrm{K}$, the mass growth rate
$\simeq 10^{10}\,\mathrm{g\,s^{-1}}$. Adopting the central density
and temperature $\rho = 160\,\mathrm{g\,cm^{-3}}$, $
T=1.5\,10^{7}\,\mathrm{K}$, as given by Allen (1973), we get for the
speed of accretion a value of about $10^{8}\,\mathrm{cm\,s^{-1}}$.
The speed of sound at the center of a solar model is
$4\,10^{7}\,\mathrm{cm\,s^{-1}}$. Its value increases to
$1.1\,10^{8}\,\mathrm{cm\;s^{-1}}$ if the more realistic central
temperature $T = T_\mathrm{eff}$ has been adopted, as mentioned
above. Hence, it seems likely that the subsonic accretion would
result in the vicinity of the BH. Of great importance that such a
low mass BH is capable to initiate the essential rearrangement of
the solar structure. The striking central temperature peak may
produce the convective core. Within the self-consistent solar model
based on a couple of sources of energy the thermonuclear burning has
decreased in importance with the trend towards smaller thermal and,
in particular, neutrino fluxes. This last presents the most
challenging question of contemporary solar physics. The availability
of a couple of sources of energy may have an immense action on
stellar evolution. For above cited example it is easy to see that
growth of the BH mass up to $M \simeq 10^{-4}M_{\odot}$ tends to
increase the solar luminosity by several times in comparison with
its present value on a time scale of $\sim 10^{9}$ years. In all
appearance, the truthful stellar evolutionary tracks may differ
markedly from those described by the present-day theory based on the
orthodox thermonuclear scenario.

\section{FINAL}

\noindent Both the capture and the incorporation of primordial black
hole is perhaps the most dramatic aspect of stellar evolution. From
this point onwards the star begins to follow another evolutionary
history than it is normally expected via the thermonuclear scenario.
Duration of this mode may extend either over billion years or may be
very transient. It is only a question of proportion between the
primary stellar mass and the initial BH mass captured. From the
condition (7) we may see that if the BH mass is above $\sim
10^{-5}M_{\odot}$, the major contribution to the luminosity comes
from the relativistic gravitational reactor. In such a case the star
evolves towards the Eddington limit. This should lead to
considerable expansion of a star and a global stability loss.
Amongst stars, which show similar physical properties, are the R
Coronae Borealis (RCB) type variables. These low-mass
hydrogen-deficient super giants  show F-G Ib spectra and
semi-regular light variations. Only between 30 and 40 of these stars
have been identified. The estimated number of RCB stars in the
Galaxy could be up $\sim$ 1000 (Lawson et al. 1990). The models
proposed for the formation of RCB stars may not maintain the
required luminosity and surface characteristics (Iben et al. 1996).
As discussed by Asplund \& Gustafsson (1996), these stars evolve
from sub-Eddington to super-Eddington luminosity.

All the above-mentioned arguments appear to reveal that many of
stars come into contact with primordial black holes during their
lifetime. Can this provoke the collapse? If so, supernovae should
occur at the rate of a few events a month. Assume the
Chapline-Hawking-Page limit on the number density of PBH with
initial masses around $10^{15}\,\mathrm{g}$. If  PBHs are clustered
in the Galaxy to the same degree as the visible matter, we obtain an
order of magnitude estimate $n_{BH}\sim 10^{9}\div
10^{10}\,\mathrm{pc^{-3}}$. Then about 100 stars of our Galaxy may
suffer fatal collisions with PBHs yearly. In the chromosphere and
upper photosphere a downward-moving PBH should produce the gamma ray
burst with the duration as short as a few seconds. The same should
occur when the upward-moving PBH escapes from the star. The total
luminosity of the bursts must have been roughly of $L\sim 10^{38}
M/M_{\odot} \; \mathrm{ erg \;s^{-1}}$, according to the condition
(7). One can envision a PBH orbiting inside the Sun. The above
considerations showed that a PBH experiences very little friction in
passing through the stellar matter. The period $P$ of revolution is
given by Kepler's third law
\begin{equation}\label{Kepler}
P=\left(\frac{3\pi}{G\overline{\rho}}\right)^{\frac{1}{2}}=1.2\,
10^{4}\,\overline{\rho}^{\; -\frac{1}{2}},
\end{equation}
where $\overline{\rho}$ is the mean density of solar matter within
the circular orbit. It is varied from $675\,\mathrm{s}$ in the
vicinity of the solar center to $167\,\mathrm{min}$ at its surface.
The latter practically coincides with the $160\,\mathrm{min}$
periodicity in radial velocity discovered by Severny, Kotov, and
Tsap (1976). One may speculate that this is the tidal forced
oscillation due to the BH companion, which is in orbital motion
inside the Sun. This is in general agreement with the existence of a
high degree of the phase coherence of oscillations over many years.

The microscopic PBHs can exist in the interior of planetary bodies
too. An important hint of such a case may be found in the Solar
system. In particular, an excess of heat flux, amounting $\sim$ 200
per cent of the solar absorbable power are observed in Jupiter and
Saturn, $\sim 4.8\,10^{11}$ and $\sim 8.6\,10^{10}\,\mathrm{MW}$,
respectively (Ingersoll et al. 1975, Reese 1971). Measurements of
the outward heat flux from another solar planets indicate excess,
amounting to a few percent only. It had been suggested that Jupiter
and Saturn could be in the core contraction stage responsible for
the production of additional internal heat flux. This raises,
however, the question of why the internal source of energy does not
exist in the case of Uranus, as with Jupiter and Saturn, whereas
their internal constitutions are closely similar. At present this
question has not yet been settled. To produce the required excess of
thermal energy on Jupiter and Saturn the masses of PBHs captured are
assumed to be reached of $\sim 4\,10^{19}$ and $\sim
7\,10^{18}\,\mathrm{g}$, respectively, according to (7). These
microscopic objects are comparable to the hydrogen atom in size
($\sim 10^{-8}\,\mathrm{cm}$). Not a single PBH, and more likely, a
swarm of them orbit freely inside the planets. These PBHs each about
of $10^{15}\,\mathrm{g}$ releases permanently about of
$5\,10^{6}\,\mathrm{MW}$ of energy, according to (7).

One can envision even a planet with the PBH acting as the
self-sufficient source of heating. Such a planet does not need the
central sun for the maintenance of animal life on its surface.  This
may last eons. The singular source of energy cannot be exhausted and
cannot die out.

\bigskip
\noindent {\bf \it Note added in proof}
\bigskip

\noindent  Rubin et al. (2001), Khlopov et al. (2002) proposed a
principally new scenario of primordial structure formation in the
models of hot Universe. These models predict phase transition in the
inflation stage period and the domain walls formation. The wall
collapse in the postinflation epoch results in the formation of
black hole clusters. As shows the results of numerical calculations,
the condition of wall existence is fulfilled for the domains with
masses exceeding $10^{15}$ g. The maximum of PBH mass distribution
falls around $10^{25}$ g and ranges up to $10^{35}$ g. The total
mass of PBH amounts to $\sim 1\%$ of the contemporary baryonic
distribution. These estimates are in harmony with studies of our
work.

\bigskip

The author thanks Prof. M. Yu. Khlopov for drawing my attention to
the results mentioned above.


\end{document}